# Non-Invasive Fuhrman Grading of Clear Cell Renal Cell Carcinoma Using Computed Tomography Radiomics Features and Machine Learning

Mostafa Nazari[1], Isaac Shiri*[2], Ghasem Hajianfar[3], Niki Oveisi[5], Hamid Abdollahi[4], Mohammad Reza Deevband[1], Mehrdad Oveisi[6]

1. Department of Medical Physics, School of Medicine, Shahid Beheshti University of Medical Sciences, Tehran, Iran.
2. Division of Nuclear Medicine and Molecular Imaging, Department of Medical Imaging, Geneva University Hospital, CH-1211 Geneva 4, Switzerland
3. Rajaie Cardiovascular Medical and Research Center, Iran University of Medical Science, Tehran, Iran
4. School of Population and Public Health, The University of British Columbia, BC, V6T 1Z4, Canada.
5. Department of Radiologic Sciences and Medical Physics, Faculty of Allied Medicine, Kerman University, Kerman, Iran
6. Department of Computer Science, University of British Columbia, Vancouver BC, Canada

**Corresponding Author:**

**Isaac Shiri**

**Address**: Division of Nuclear Medicine and Molecular Imaging,
Department of Medical Imaging,
Geneva University Hospital,
CH-1211 Geneva 4, Switzerland

**Email**: Isaac.Shiri@etu.unige.ch

**Tel**: +41766269359

**Abstract**

**Purpose:** To identify optimal classification methods for computed tomography (CT) radiomics-based preoperative prediction of clear cells renal cell carcinoma (ccRCC) grade.

**Methods and material:** Seventy-one ccRCC patients (31 low-grade and 40 high-grade) were included in the study. All tumors were segmented manually on CT images, and three image preprocessing techniques (Laplacian of Gaussian, wavelet filter, and discretization of the intensity values) were applied on tumor volumes. In total, 2530 radiomics features (tumor shape and size, intensity statistics, and texture) were extracted from each segmented tumor volume. Univariate analysis was performed to assess the association of each feature with the histological condition. In the case of multivariate analysis, the following was implemented: three feature selection including the least absolute shrinkage and selection operator (LASSO), student's t-test and minimum Redundancy Maximum Relevance (mRMR) algorithms. These selected features were then used to construct three classification models (SVM, random forest, and logistic regression) to discriminate the high from low-grade ccRCC at nephrectomy. Lastly, multivariate model performance was evaluated on the bootstrapped validation cohort using the area under receiver operating characteristic curve (AUC).

**Results:** Univariate analysis demonstrated that among different image sets, 128 bin discretized images have statistically significant different (q-value $< 0.05$) texture parameters with a mean of AUC $0.74 \pm 3$ (q-value $< 0.05$). The three ML-based classifier shows proficient discrimination of the high from low-grade ccRCC. The AUC was 0.78 in logistic regression, 0.62 in random forest, and 0.83 in SVM model, respectively.

**Conclusion:** Radiomics features can be a useful and promising non-invasive method for preoperative evaluation of ccRCC Fuhrman grades.

## Introduction

Renal cell carcinoma (RCC) is the seventh most common cancer in the world, with a mortality rate of 140,000 per year (1). The most common types of renal cancer cells consists of clear cells RCC (ccRCC), papillary RCC (pRCC), and chromophobe RCC (chRCC) (2, 3). Approximately 70% of kidney cancers are made up of ccRCC, pRCC accounts for 10-15% of kidney cancers, and chRCC is the least common type with only 5% of kidney cancer cases (4). A ccRCC diagnosis has a survival rate of less than 5-years and a higher risk of metastasis compared to pRCC and chRCC (5).

The most important step for a physician during cancer diagnosis and treatment is tumor staging and grading. Tumor grading is a description of the differentiation of tumor tissue cells relative to normal tissue cells. It is an indicator of how quickly a tumor is expected to grow and spread. The Fuhrman grading system is widely accepted (6) and is based on the assessment of the following cell nucleus characteristics: nuclear size, nuclear shape, and nucleolar prominence. Based on these assessments, the tumor will be classified into one of four different grades (I-IV). Grades I and II are considered as low-grade tumors with a favorable prognosis, while grades III and IV accounted for high-grade tumors and have an unfavorable prognosis (7).

Currently, fine needle aspiration (FNA) and imaging-guided biopsies are the gold standard methods for preoperative kidney tumor grading. However, these techniques have some drawbacks including infection, bleeding, tumor cells spreading, and provide limited information regarding the whole tumor. For example, a pathologist may diagnose a ductal carcinoma in situ (DCIS) as a non-invasive breast cancer based on the FNA breast sample obtained, when in fact, the patient has infiltrating ductal carcinoma (IDC) in a nearby area (8, 9).

In the recent years, several new non-invasive therapeutic methods for RCC have been developed, including radiofrequency ablation, cryoablation, and active surveillance(10). However, there lacks  proper criterion for patient management with these non-invasive/minimally invasive treatment methods, and are often treated surgically post-diagnosis (11, 12). Therefore, it is desirable to produce individualized treatment strategies, where radical approaches (e.g. surgery) are taken for aggressive or high-grade tumors (III, IV) ccRCC, and conservative management (e.g. active surveillance) is provided for low-grade (I, II) lesions (13). To guide this decision, a non-invasive and accurate method for Fuhrman grading of renal cell carcinoma tumors preoperatively is desirable. For this purpose, two approaches are attractive for researchers at present. One is the Apparent diffusion coefficient (ADC) value in MRI imaging (14), and the other is CT-based semi-quantitative and quantitative techniques (15, 16).

Radiomics refers to the comprehensive quantification of tumor phenotypes in order to uncover disease characteristics that fail to be explained by the naked eye (17-20). In fact, it can be said that radiomics serves as the bridge between medical imaging and personalized medicine (21). Radiomics is a new era of science which faces many challenges, including image acquisition (22), reconstruction (23, 24), processing (25), and model development(26, 27) to provide robust and reproducible models. Previous studies have shown that the radiomics signature is valuable for differentiating high/low grade ccRCC tumors (28, 29). This study aims to construct a radiomics feature–based machine learning model to predict the Fuhrman Grade of ccRCC patients preoperatively.

## Methods and Materials

Flow chart of the current study is illustrated in Figure 1

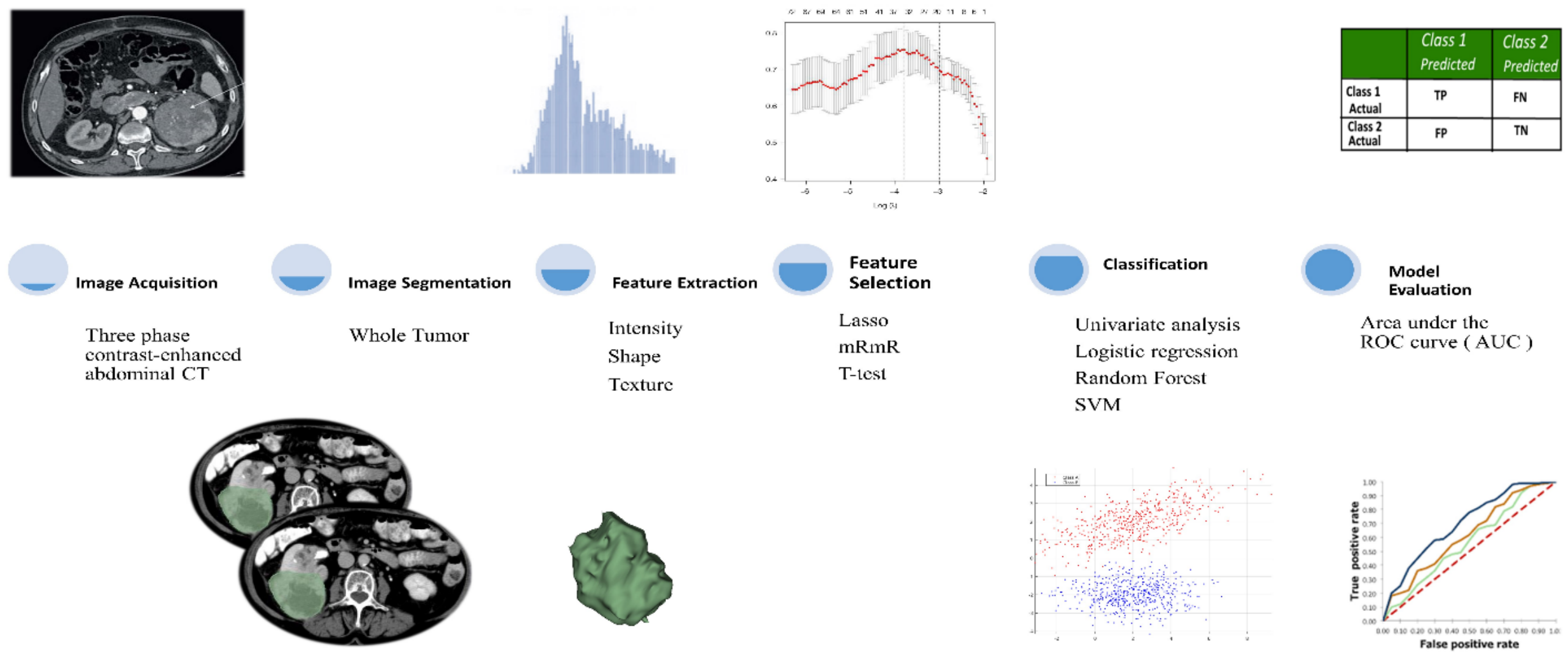

**Figure 1**. Illustrates the process flow followed in the paper.

- **Data**

Data was collected from cancer image archive databases from 1980 to 2016 (30). (Table 1)

**Table 1**.Clinical characteristics of Clear Cell Renal Cell Carcinoma

| **Characteristic** | |
|---|---|
| | **Patient (N=71)** |
| **Gender** | |
| **Male** | **51 (71%)** |
| **Female** | **20 (29%)** |
| **Age, Y** | **60.3 ± 11.7** |
| **Grade** | |
| **Low Grade (I ,II)** | **31** |
| **High Grade (III,IV)** | **40** |

- **Image acquisition technique**

All patients had undergone a three phasic CT scan, including 1) a routine unenhanced CT scan, 2) a corticomedullary phase (CMP) contrast enhanced scan starting 40 seconds after the contrast material injection, and 3) a nephrographic phase (NP) contrast-enhanced scan performed 70–90 seconds after intravenous injection of iodinated contrast material. The iodine content (300 mg/mL) was infused at an infusion rate of 3 mL/s at an infusion dose of 80–100 mL. All subjects were scanned using a GE and Siemens CT machine with a tube voltage of 120 kV and a tube current of 150–300 mA with daily clinical reconstruction parameters.

- **Tumor Segmentation**

In this study manual volume of interest (VOI) segmentation was performed and verified by a radiologist with 3D slicers.

- **Image pre-processing**

Prior to feature extraction, the voxel size resampling method was applied on the images to create an isotropic dataset. This allowed comparisons between image data from different samples and scanners (31). Laplacian of Gaussian (LOG), wavelet decomposition (WAV), and discretization into 32, 64, and 128 bins preprocessing was performed to generate different set of features. For the LOG filter, different sigma values were used to extract fine, medium, and coarse features. Wavelet filtering yields 8 decompositions per level: all possible combinations of applying either a High or a Low pass filter in each of the three dimensions including HHH, HHL, HLH, HLL, LHH, LHL, LLH, and LLL. Preprocessing steps (including discretization, LOG, and wavelet) were also performed on all intensity, histogram, and textural features.

- **Feature extraction**

Radiomics features were extracted through the PyRadiomics open source python library. Extracted features were then categorized into the following subgroups. Firstly, shape features depict the shape of the tumor volume (VOI) and geometric properties such as volume, maximum surface, tumor compactness, and sphericity. Furthermore, first-order statistics features describe the distribution of voxel intensities within tumor volumes. This includes mean, median, maximum, and minimum values of the voxel intensities on the image. Second-order statistics features (known as textural features) are used as a method to measure inter-relationships between voxel distributions in tumor volumes. This is reflective of changes in image space gray levels. These feature groups include: gray-level co-occurrence matrix (GLCM), gray-level run-length matrix (GLRLM), gray-level size-zone matrix (GLSZM), and gray-level dependence matrix (GLDM) features (see Table 2).

**Table 2**.Radiomics Features

| • **First Order Statistics (FOS)** | • **Gray Level Co-occurrence Matrix (GLCM)** | • **Gray Level Run Length Matrix (GLRLM)** |
|---|---|---|
| •<br>• Energy<br>• Total Energy<br>• Entropy<br>• Minimum<br>• 10th percentile<br>• 90th percentile<br>• Maximum<br>• Mean<br>• Median<br>• Interquartile Range<br>• Range<br>• Mean Absolute Deviation (MAD)<br>• Robust Mean Absolute Deviation (rMAD)<br>• Root Mean Squared (RMS)<br>• Standard Deviation<br>• Skewness<br>• Kurtosis<br>• Variance<br>• 19. Uniformity | •<br>• Autocorrelation<br>• Joint Average<br>• Cluster Prominence<br>• Cluster Shade<br>• Cluster Tendency<br>• Contrast<br>• Correlation<br>• Difference Average<br>• Difference Entropy<br>• Difference Variance<br>• Joint Energy<br>• Joint Entropy<br>• Informal Measure of Correlation (IMC) 1<br>• Informal Measure of Correlation (IMC) 2<br>• Inverse Difference Moment (IDM)<br>• Inverse Difference Moment Normalized (IDMN)<br>• Inverse Difference (ID)<br>• Inverse Difference Normalized (IDN)<br>• Inverse Variance<br>• Maximum Probability<br>• Sum Average<br>• Sum Entropy<br>• 23. Sum of Squares | •<br>• Short Run Emphasis (SRE)<br>• Long Run Emphasis (LRE)<br>• Gray Level Non-Uniformity (GLN)<br>• Gray Level Non-Uniformity Normalized (GLNN)<br>• Run Length Non-Uniformity (RLN)<br>• Run Length Non-Uniformity Normalized (RLNN)<br>• Run Percentage (RP)<br>• Gray Level Variance (GLV)<br>• Run Variance (RV)<br>• Run Entropy (RE)<br>• Low Gray Level Run Emphasis (LGLRE)<br>• High Gray Level Run Emphasis (HGLRE)<br>• Short Run Low Gray Level Emphasis (SRLGLE)<br>• Short Run High Gray Level Emphasis (SRHGLE)<br>• Long Run Low Gray Level Emphasis (LRLGLE)<br>• 16. Long Run High Gray Level Emphasis (LRHGLE) |
| • **Shape Features** | • **Gray Level Size Zone Matrix (GLSZM)** | • **Gray Level Dependence Matrix (GLDM)** |
| •<br>• Volume<br>• Surface Area<br>• Surface Area to Volume ratio<br>• Sphericity<br>• Spherical Disproportion<br>• Maximum 3D diameter<br>• Maximum 2D diameter (Slice)<br>• Maximum 2D diameter (Column)<br>• Maximum 2D diameter (Row)<br>• Major Axis<br>• Minor Axis<br>• Least Axis<br>• Elongation<br>• Flatness | •<br>• Small Area Emphasis (SAE)<br>• Large Area Emphasis (LAE)<br>• Gray Level Non-Uniformity (GLN)<br>• Gray Level Non-Uniformity Normalized (GLNN)<br>• Size-Zone Non-Uniformity (SZN)<br>• Size-Zone Non-Uniformity Normalized (SZNN)<br>• Zone Percentage (ZP)<br>• Gray Level Variance (GLV)<br>• Zone Variance (ZV)<br>• Zone Entropy (ZE)<br>• Low Gray Level Zone Emphasis (LGLZE)<br>• High Gray Level Zone Emphasis (HGLZE)<br>• Small Area Low Gray Level Emphasis (SALGLE)<br>• Small Area High Gray Level Emphasis (SAHGLE)<br>• Large Area Low Gray Level Emphasis (LALGLE)<br>• 16. Large Area High Gray Level Emphasis (LAHGLE) | •<br>• Small Dependence Emphasis (SDE)<br>• Large Dependence Emphasis (LDE)<br>• Gray Level Non-Uniformity (GLN)<br>• Dependence Non-Uniformity (DN)<br>• Dependence Non-Uniformity Normalized (DNN)<br>• Gray Level Variance (GLV)<br>• Dependence Variance (DV)<br>• Dependence Entropy (DE)<br>• Low Gray Level Emphasis (LGLE)<br>• High Gray Level Emphasis (HGLE)<br>• Small Dependence Low Gray Level Emphasis (SDLGLE)<br>• Small Dependence High Gray Level Emphasis (SDHGLE)<br>• Large Dependence Low Gray Level Emphasis (LDLGLE)<br>• 14. Large Dependence High Gray Level Emphasis (LDHGLE)<br>• **Neighboring Gray Tone Difference Matrix (NGTDM)**<br>•<br>• 1-Coarseness<br>• 2-Contrast<br>• 3-Busyness<br>• 4-Complexity<br>• 5- Strength |

- **Univariate analysis**

For univariate analysis, early Pearson correlation tests between features were used to eliminate highly correlated features. Student's t-tests were then used for comparisons between two groups. To control for the False Discovery Rate (FDR) in multiple hypothesis testing, the Benjamini-Hochberg (FDR) correction method was applied on p-values, and the ultimately reported q-value (32).

- **Feature Set preprocessing**

Due to the different ranges of radiomics features, without feature normalization, some features might appear as a larger weight, while others might appear as a lower weight. This depends on the distribution of feature values. To eradicate this, z-score normalization was applied to the feature values (33).

- **Feature Selection**

Three different feature selections methods (Table 3) were implemented in this framework: enhanced variable selection algorithms based on the least absolute shrinkage and selection operator methods (34), student's t-test (26, 27), and MRMR (Minimum Redundancy Maximum Relevance) algorithm.

- **Multivariate Machine Learning Classifier**

The following three classifiers (Table 3) were implemented and compared: logistic regression, random forest, and support vector machines (SVM).

- **Model evaluation**

The cross validation (CV) technique was applied in order to tune the model parameters. Furthermore, bootstrapped datasets were used for model evaluations. The predictive power of all models was investigated using the area under the receiver operator characteristic (ROC) curve (AUC). All analysis and evaluation were performed by using the programming software R (version 3.5.2).

**Table 3.** Feature selection and Classification methods

| Feature Selection Methods | Abbreviation | Classification Methods | Abbreviation |
|---|---|---|---|
| T student test | | Logistic Regression | LR |
| Minimum Redundancy Maximum Relevance | MRMR | Random Forest | RF |
| least absolute shrinkage and selection operator | LASSO | Support Vector Machine | SVM |

**Results**

After applying inclusion/exclusion criteria, 71 (31 low-grade and 40 high-grade) patients were selected. The mean ages of the low- and high-grade groups were 60.05 and 60.08 years old, respectively. In total, there are 51 male and 20 female participants.

Univariate analysis demonstrated that among filtered and non-filtered images only the 128 bin discretized images have a statistically significant difference (q-value < 0.05) in texture parameters with a mean AUC of 0.74±3 (q-value < 0.05). These features include Long Run High Gray Level Emphasis from GLRLM (AUC: 77, q-value: 0.0002), Cluster Tendency from GLCM (AUC: 72, q-value: 0.001), Contrast from NGTDM (AUC: 74, q-value: 0.03), and Dependence Non-Uniformity from GLDM (AUC: 72, q-value: 0.04) (Figure 2).

Table 4 shows the AUC (95% CI) of three different ML-based classifiers. As shown in the table, there is a wide performance range from 0.5 to 0.86. Three different feature selection methods were applied prior to the implementation of each ML-based classifier to determine which is the best for that model. The results demonstrated that the lasso method performs the best for logistic regression. Furthermore, the student's t-test proved to be the best for random forest and SVM classifier models. Results for logistic regression suggested that 128 bin discretized images and fine LoG features have the highest performance with a mean of AUC 0.75. According to the results, the predictive performance of the random forest model has a range of 0.48 to 0.67. Among these, wavelet filtered images showed lowest performance and 128 bin discretized images showed highest performance. Among the three classifiers, SVM with a student's t-test feature selection presented the best predictive performance. SVM with Coarse LoG features demonstrated a mean AUC 0.83 (Figure 2).

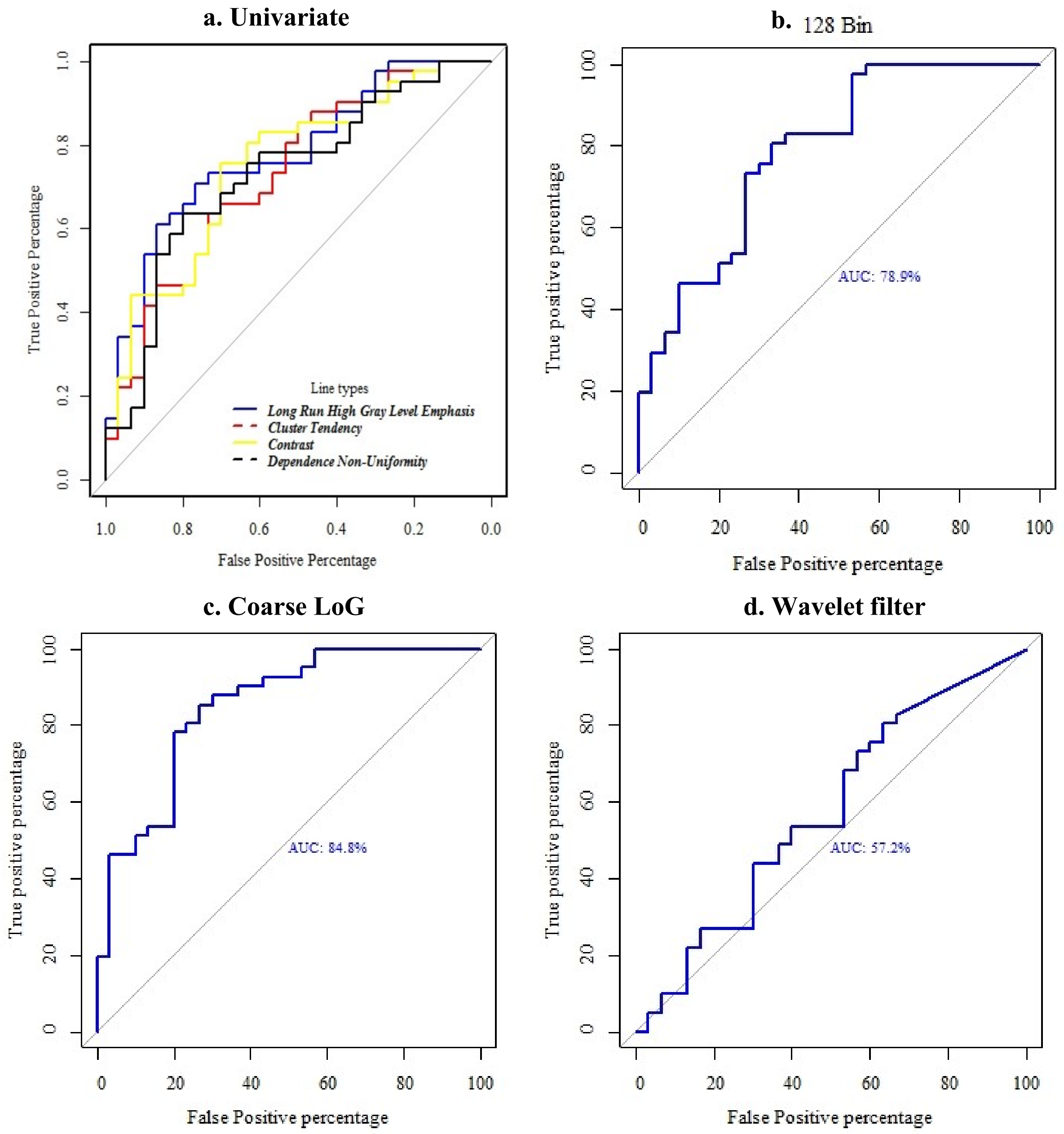

**Figure 2**. AUC for discrimination of the high from low grade ccRCC. a. Univariate analysis of best predictor, b. LR model with 128 bin discretizing, c. SVM model with Coarse LoG filter, d. RF model with Wavelet filter. AUC: Area under receiver operating characteristic curve, LR: logistic regression, SVM: Support Vector Machine, RF: Random Forest.

**Table 3.** Classifiers performance with different pre-processing

| | AUC (95%CI) | | |
|---|---|---|---|
| | **LR** | **RF** | **SVM** |
| **Original image** | 0.68 ±0.08 | 0.62 ±0.07 | 0.76 ±0.07 |
| **32_bin** | 0.73 ±0.09 | 0.56 ±0.06 | 0.70 ±0.08 |
| **64 _bin** | 0.70 ±0.07 | 0.55 ±0.06 | 0.65 ±0.07 |
| **128_bin** | 0.75 ±0.08 | 0.60 ±0.08 | 0.77 ±0.08 |
| **LoG** | 0.69 ±0.08 | 0.53 ±0.06 | 0.62 ±0.06 |
| **LoG_ sigma.0.5** | 0.74 ±0.11 | 0.53 ±0.02 | 0.64 ±0.04 |
| **LoG_sigma.1.0** | 0.65 ±0.09 | 0.56 ±0.02 | 0.72 ±0.07 |
| **LoG_sigma.1.5** | 0.68 ±0.09 | 0.55 ±0.06 | 0.74 ±0.08 |
| **LoG_sigma.2.0** | 0.73 ±0.11 | 0.54 ±0.05 | 0.76 ±0.07 |
| **LoG_sigma.2.5** | 0.74 ±0.10 | 0.55 ±0.09 | 0.77 ±0.06 |
| **LoG_sigma.3.0** | 0.62 ±0.06 | 0.57 ±0.04 | 0.79 ±0.08 |
| **LoG_sigma.3.5** | 0.65 ±0.09 | 0.60 ±0.12 | 0.81 ±0.06 |
| **LoG_sigma.4.0** | 0.62 ±0.06 | 0.62 ±0.06 | 0.83 ±0.08 |
| **LoG_sigma.4.5** | 0.70 ±0.08 | 0.56 ±0.05 | 0.78 ±0.06 |
| **LoG_sigma.5.0** | | | |
| **WAVELET** | 0.62 ±0.05 | 0.58 ±0.04 | 0.71 ±0.07 |
| **Wav_HHL** | 0.65 ±0.06 | 0.55 ±0.08 | 0.62 ±0.06 |
| **Wav_HLH** | 0.67 ±0.08 | 0.53 ±0.03 | 0.65 ±0.07 |
| **Wav_LHH** | 0.62 ±0.06 | 0.57 ±0.05 | 0.76 ±0.08 |
| **Wav_HLL** | 0.68 ±0.07 | 0.56 ±0.06 | 0.59 ±0.08 |
| **Wav_LLH** | 0.62 ±0.06 | 0.53 ±0.05 | 0.65 ±0.08 |
| **Wav_LHL** | 0.62 ±0.06 | 0.55 ±0.08 | 0.75 ±0.10 |
| **Wav_LLL** | | | |

## Discussion

There is an important association between the Fuhrman grade and patient's prognosis. There are several non-invasive methods proposed to predict the ccRCC Fuhrman grade preoperatively. In MR imaging, the apparent diffusion coefficient (ADC) value is known to be an indicator of tumor activity. Several studies have assessed the utility of the apparent diffusion coefficient (ADC) in distinguishing low- and high-grade clear cell RCC (14, 35). These studies showed that magnetic resonance imaging (MRI) has an acceptable predictive accuracy in the preoperative detection of the high grade RCC (AUC was 0.80) (36). However, MRI is a costly process and a wide range of ADC values for ccRCC have been reported in the literature (37, 38). Therefore, their repeatability needs to be validated further. Furthermore, a large number of CT-based semi-quantitative and quantitative studies have attempted to classify low- and high-grade ccRCC (15, 16). These studies showed that CT is a promising method for this work.

Radiomics approach convert medical images into quantitative, high-dimensional, and mineable features to predict tumor status. However, the abundance of predictive modeling techniques means that it is important to choose the correct one for predicting tumor status. As some previous radiomics studies (16, 28), for Fuhrman grade prediction did not include shape features in their analyses, this study combined shape features and texture features to differentiate low- and high-grades of ccRCC. It was observed that shape features cannot be ignored from multivariate machine learning models.

Univariate analysis of extracted radiomics features demonstrated that among filtered and non-filtered images only the 128 bin discretized images showed statistically significant texture parameters. In a similar univariate analysis, Zhan Feng *et al* (29) analyzed CT texture parameters and found effective quantitative parameters to evaluate the heterogeneity of ccRCC. After applying the LoG filter, they reported that only entropy has a statistically significant difference after FDR corrections in all image phases. In this study four features showed statistically significant differences between two groups. These features include: Long Run High Gray Level Emphasis from GLRLM, Cluster Tendency from GLCM Contrast from NGTDM, and Dependence Non-Uniformity from GLDM matrix. Among these features, the Long Run High Gray Level Emphasis demonstrated the highest AUC (AUC: 77, q-value: 0.0002).

The first machine learning model applied in this study was logistic regression. It is a machine learning classification algorithm used to predict the class probability of a categorical dependent variable. It was observed that among three different features selection methods, the best results for the Logistic Regression model was obtained when using the lasso algorithm. These results suggest that the AUC logistic regression model is approximately similar to results obtained in previous studies. Juice Dinga *et al* (16) used a texture-score based logistic regression model on a training cohort and the AUC was 0.878. When predictive models were applied on the validation cohort, good results were still obtained (AUC > 0.670). Jun Shu1 *et al* (39) extracted radiomics features from the corticomedullary (CMP) and nephrographic phases (NP) of CT images of all patients. They constructed logistic regression classification models to discriminate high and low grades ccRCC. Application of the model on CMP and NP showed an AUC of 0.766 (95% CI:0.709-0.816) and 0.818 (95% CI:0.765-0.838), respectively. Another machine learning model applied was random forest, which is an ensemble learning method that consists of a collection of decision trees. It uses a weighted average of those trees for the final decision (40). It works correctly for a large range of data, but is susceptible to overfitting. In this study, applying the random forest model on the dataset yielded unsatisfactory results (table 1). The SVM was the best-performing classifier in this study. SVM creates a decision boundary between two classes that enables the prediction of labels from one or more feature vectors. After applying the SVM model on filtered and unfiltered images, the best classification result was obtained when coarse LoG features were used with a mean AUC of 0.81. LoG filtering is an advanced image-filtering method that combines Laplacian filtering and Gaussian filtering. In a similar single-center retrospective study (28) , the performance of quantitative CT texture analysis combined with different ML based classifier methods is evaluated for discriminating low and high grade ccRCC. Despite differences in procedure, they also determined that highest predictive performance is achieved by an SVM classifier. In summary, both of these studies support each other with a similar conclusion that CT texture analysis is a useful and promising non-invasive method to predict the Fuhrman grades of ccRCCs preoperatively.

The limitations of this study were as follows. (1) This study was a retrospective study with no external data validation, (2) the sample size was relatively small, (3) since the tumor boundary is manually drawn, the interference of the volume effect cannot be completely avoided.

**Conclusion**

The results of this study show that CT based SVM classifier with t-test features selection can be a useful and promising non-invasive method for the prediction of low and high Fuhrman nuclear grade ccRCCs. Additionally, the results demonstrated that 128 bin discretized pre-processing is an effective method under these conditions. Large, multicenter and externally prospective studies are needed for further validation of CT base machine learning models.

## Reference

1. Capitanio U, Montorsi F. Renal cancer. The Lancet. 2016;387(10021):894-906.
2. Srigley JR, Delahunt B, Eble JN, Egevad L, Epstein JI, Grignon D, et al. The International Society of Urological Pathology (ISUP) vancouver classification of renal neoplasia. The American journal of surgical pathology. 2013;37(10):1469-89.
3. Marconi L, Dabestani S, Lam TB, Hofmann F, Stewart F, Norrie J, et al. Systematic review and meta-analysis of diagnostic accuracy of percutaneous renal tumour biopsy. European urology. 2016;69(4):660-73.
4. Muglia VF, Prando A. Renal cell carcinoma: histological classification and correlation with imaging findings. Radiologia brasileira. 2015;48(3):166-74.
5. Delahunt B, Cheville JC, Martignoni G, Humphrey PA, Magi-Galluzzi C, McKenney J, et al. The International Society of Urological Pathology (ISUP) grading system for renal cell carcinoma and other prognostic parameters. The American journal of surgical pathology. 2013;37(10):1490-504.
6. Lohse CM, Blute ML, Zincke H, Weaver AL, Cheville JC. Comparison of standardized and nonstandardized nuclear grade of renal cell carcinoma to predict outcome among 2,042 patients. American journal of clinical pathology. 2002;118(6):877-86.
7. Delahunt B. Advances and controversies in grading and staging of renal cell carcinoma. Modern Pathology. 2009;22(S2):S24.
8. Lechevallier E, André M, Barriol D, Daniel L, Eghazarian C, De Fromont M, et al. Fine-needle percutaneous biopsy of renal masses with helical CT guidance. Radiology. 2000;216(2):506-10.
9. Al Nazer M, Mourad WA. Successful grading of renal-cell carcinoma in fine-needle aspirates. Diagnostic cytopathology. 2000;22(4):223-6.
10. Kutikov A, Kunkle DA, Uzzo RG. Focal therapy for kidney cancer: a systematic review. Current opinion in urology. 2009;19(2):148-53.
11. Volpe A, Panzarella T, Rendon RA, Haider MA, Kondylis FI, Jewett MA. The natural history of incidentally detected small renal masses. Cancer. 2004;100(4):738-45.
12. Bratslavsky G, Kirkali Z. The changing face of renal-cell carcinoma. Journal of endourology. 2010;24(5):753-7.
13. Kunkle DA, Egleston BL, Uzzo RG. Excise, ablate or observe: the small renal mass dilemma—a meta-analysis and review. The Journal of urology. 2008;179(4):1227-34.
14. Yoshida R, Yoshizako T, Hisatoshi A, Mori H, Tamaki Y, Ishikawa N, et al. The additional utility of apparent diffusion coefficient values of clear-cell renal cell carcinoma for predicting metastasis during clinical staging. Acta radiologica open. 2017;6(1):2058460116687174.
15. Ishigami K, Leite LV, Pakalniskis MG, Lee DK, Holanda DG, Kuehn DM. Tumor grade of clear cell renal cell carcinoma assessed by contrast-enhanced computed tomography. SpringerPlus. 2014;3(1):694.
16. Ding J, Xing Z, Jiang Z, Chen J, Pan L, Qiu J, et al. CT-based radiomic model predicts high grade of clear cell renal cell carcinoma. European journal of radiology. 2018;103:51-6.
17. Shiri I, Maleki H, Hajianfar G, Abdollahi H, Ashrafinia S, Hatt M, et al. Next Generation Radiogenomics Sequencing for Prediction of EGFR and KRAS Mutation Status in NSCLC Patients Using Multimodal Imaging and Machine Learning Approaches. arXiv preprint arXiv:190702121. 2019.
18. Hajianfar G, Shiri I, Maleki H, Oveisi N, Haghparast A, Abdollahi H, et al. Non-Invasive MGMT Status Prediction in GBM Cancer Using Magnetic Resonance Images Radiomics Features: Univariate and Multivariate Radiogenomics Analysis. World Neurosurgery. 2019.

19. Shiri I, Maleki H, Hajianfar G, Abdollahi H, Ashrafinia S, Oghli MG, et al., editors. PET/CT Radiomic Sequencer for Prediction of EGFR and KRAS Mutation Status in NSCLC Patients. 2018 IEEE Nuclear Science Symposium and Medical Imaging Conference Proceedings (NSS/MIC); 2018: IEEE.
20. Abdollahi H, Mahdavi SR, Shiri I, Mofid B, Bakhshandeh M, Rahmani K. Magnetic resonance imaging radiomic feature analysis of radiation-induced femoral head changes in prostate cancer radiotherapy. Journal of cancer research and therapeutics. 2019;15(8):11.
21. Lambin P, Leijenaar RT, Deist TM, Peerlings J, De Jong EE, Van Timmeren J, et al. Radiomics: the bridge between medical imaging and personalized medicine. Nature Reviews Clinical Oncology. 2017;14(12):749.
22. Abdollahi H, Shiri I, Heydari M. Medical Imaging Technologists in Radiomics Era: An Alice in Wonderland Problem. Iranian journal of public health. 2019;48(1):184.
23. Shiri I, Rahmim A, Ghaffarian P, Geramifar P, Abdollahi H, Bitarafan-Rajabi A. The impact of image reconstruction settings on 18F-FDG PET radiomic features: multi-scanner phantom and patient studies. European radiology. 2017;27(11):4498-509.
24. Shiri I, Ghafarian P, Geramifar P, Leung KH-Y, Ghelichoghli M, Oveisi M, et al. Direct attenuation correction of brain PET images using only emission data via a deep convolutional encoder-decoder (Deep-DAC). European radiology. 2019:1-13.
25. Shiri I, Abdollahi H, Shaysteh S, Mahdavi SR. Test-retest reproducibility and robustness analysis of recurrent glioblastoma MRI radiomics texture features. Iranian Journal of Radiology. 2017(5).
26. Abdollahi H, Mofid B, Shiri I, Razzaghdoust A, Saadipoor A, Mahdavi A, et al. Machine learning-based radiomic models to predict intensity-modulated radiation therapy response, Gleason score and stage in prostate cancer. La radiologia medica. 2019;124(6):555-67.
27. Abdollahi H, Mostafaei S, Cheraghi S, Shiri I, Mahdavi SR, Kazemnejad A. Cochlea CT radiomics predicts chemoradiotherapy induced sensorineural hearing loss in head and neck cancer patients: a machine learning and multi-variable modelling study. Physica Medica. 2018;45:192-7.
28. Bektas CT, Kocak B, Yardimci AH, Turkcanoglu MH, Yucetas U, Koca SB, et al. Clear cell renal cell carcinoma: machine learning-based quantitative computed tomography texture analysis for prediction of Fuhrman nuclear grade. European radiology. 2019;29(3):1153-63.
29. Feng Z, Shen Q, Li Y, Hu Z. CT texture analysis: a potential tool for predicting the Fuhrman grade of clear-cell renal carcinoma. Cancer Imaging. 2019;19(1):6.
30. Akin O, Elnajjar P, Heller M, Jarosz R, Erickson B, Kirk S, et al. Radiology data from the cancer genome atlas kidney renal clear cell carcinoma [TCGA-KIRC] collection. The Cancer Imaging Archive. 2016.
31. Shafiq-ul-Hassan M, Zhang GG, Latifi K, Ullah G, Hunt DC, Balagurunathan Y, et al. Intrinsic dependencies of CT radiomic features on voxel size and number of gray levels. Medical physics. 2017;44(3):1050-62.
32. Haynes W. Benjamini–hochberg method. Encyclopedia of systems biology. 2013:78-.
33. Kickingereder P, Götz M, Muschelli J, Wick A, Neuberger U, Shinohara RT, et al. Large-scale radiomic profiling of recurrent glioblastoma identifies an imaging predictor for stratifying anti-angiogenic treatment response. Clinical Cancer Research. 2016;22(23):5765-71.
34. Guo P, Zeng F, Hu X, Zhang D, Zhu S, Deng Y, et al. Improved variable selection algorithm using a LASSO-type penalty, with an application to assessing hepatitis B infection relevant factors in community residents. PloS one. 2015;10(7):e0134151.
35. Rosenkrantz AB, Niver BE, Fitzgerald EF, Babb JS, Chandarana H, Melamed J. Utility of the apparent diffusion coefficient for distinguishing clear cell renal cell carcinoma of low and high nuclear grade. American Journal of Roentgenology. 2010;195(5):W344-W51.

36. Maruyama M, Yoshizako T, Uchida K, Araki H, Tamaki Y, Ishikawa N, et al. Comparison of utility of tumor size and apparent diffusion coefficient for differentiation of low-and high-grade clear-cell renal cell carcinoma. Acta Radiologica. 2015;56(2):250-6.
37. Zhang J, Mazaheri Tehrani Y, Wang L, Ishill NM, Schwartz LH, Hricak H. Renal masses: characterization with diffusion-weighted MR imaging—a preliminary experience. Radiology. 2008;247(2):458-64.
38. Goyal A, Sharma R, Bhalla AS, Gamanagatti S, Seth A, Iyer VK, et al. Diffusion-weighted MRI in renal cell carcinoma: a surrogate marker for predicting nuclear grade and histological subtype. Acta Radiologica. 2012;53(3):349-58.
39. Shu J, Tang Y, Cui J, Yang R, Meng X, Cai Z, et al. Clear cell renal cell carcinoma: CT-based radiomics features for the prediction of Fuhrman grade. European journal of radiology. 2018;109:8-12.
40. Breiman L. Random Forests, Vol. 45. Mach Learn. 2001;1.